\documentclass[conference]{IEEEtran}
\IEEEoverridecommandlockouts
\usepackage{cite}
\usepackage{amsmath,amssymb,amsfonts}
\usepackage{graphicx}
\usepackage{textcomp}
\usepackage{xcolor}

\usepackage{multirow}
\usepackage{tabularx}
\usepackage{algorithm}
\usepackage{algpseudocode}
\usepackage{stfloats}
\usepackage{url}

\def\BibTeX{{\rm B\kern-.05em{\sc i\kern-.025em b}\kern-.08em
    T\kern-.1667em\lower.7ex\hbox{E}\kern-.125emX}}

\begin{document}


\title{Component Benchmark: Hierarchical Model Profiling for Large-scale Recommendation Systems}

\author{
\IEEEauthorblockN{Dharak Kharod\textsuperscript{*}, Yuzhen Huang\textsuperscript{*},
Zhou Wang, Jackie Xu, Fuzail Khan, Jacky Zhou, Hao Yan, Lidong Zhao,\\
Xizhou Feng, Yvonne Liu, Karthik Jayaraman, Praveen Ramachandran,
Vishwa Karia, Yashasvi Makin}
\IEEEauthorblockA{\textit{Meta Platforms Inc}, Sunnyvale, California, USA\\
\{dharakk, yuzhenhuang, zhouwang, jackiexu0313, fuzailkhan, junqingz, haoyan17, lidong, fengx\}@meta.com\\
\{yvliu, karthikjay, pramacha, vishwakaria, yashasvi\}@meta.com}
\thanks{\textsuperscript{*}Equal contribution.}
}

\maketitle

\begin{abstract}
Large-scale recommendation models pose distinct, under-explored profiling challenges. Most recommendation model architectures are structurally heterogeneous, intermixing memory-bandwidth-bound operations, small compute-bound dense layers, dynamic shapes from jagged categorical features, and low-arithmetic-intensity operations. Recommendation models evolve rapidly as modeling engineers experiment with compositions, often written without visibility into hardware execution characteristics. Standard profiling tools offer either end-to-end throughput or operator-level traces, but cannot attribute performance to the submodules that practitioners reason about. We present Component Benchmark (CB), a profiling system that independently characterizes each submodule performance in a hierarchical manner, providing a tree-structured, interactive visualization that brings performance clarity to ML practitioners. At its core, CB provides a simple yet extensible, submodule-based benchmarking framework with a plugin architecture that enables hierarchical performance analysis. These large-scale recommendation models are TB-scale, run on thousands of GPUs and ingest 100B examples per day. We demonstrate CB’s effectiveness on common open sourced models and discuss how CB has been leveraged to accelerate modern recommendation model performance analysis and optimization.
\end{abstract}


\section{Introduction}

Recent advances in AI/ML have been largely driven by scaling laws that push both model size and data volume to unprecedented levels, requiring efficient training and inference on modern GPUs or accelerators. However, performance optimization and bottleneck analysis in modern ML systems have become increasingly complex. This is mainly due to diverse and rapidly evolving models/datasets, heterogeneous training platforms, and high-volume profiling and telemetry data.

Model architecture innovation is one of the most active areas in AI/ML research. Large language models have converged to a few representative modeling backbones (e.g., Transformer~\cite{vaswani2017attention}); however, models for recommendation systems have not yet converged, and this remains an active research area. Modern recommendation systems have evolved from collaborative filtering / matrix factorization~\cite{koren2009matrix} to advanced neural networks with architectures like FM~\cite{rendle2010factorization}, DCN~\cite{wang2017deep}, DHEN~\cite{zhang2022dhen}, Interformer~\cite{zeng2024interformer}, and HSTU~\cite{zhai2024actions}. Within these architectures, there are inefficiencies stemming from small operator dimensions, redundant compute, repetitive kernels, etc., requiring a generic performance profiling system to analyze the bottlenecks.



Model architecture alone is insufficient for accurate performance profiling; the characteristics of training and inference data also play a critical role. For example, input sparsity in categorical features can significantly affect the performance of recommendation models, while sequence length distributions have a major impact on the efficiency of large language models. Meaningful profiling, therefore, requires the use of real datasets. In addition, models and data are typically embedded within complex training and inference stacks and controlled through numerous configuration options. A practical profiling system must be platform-agnostic and capable of handling models and datasets across multiple ML frameworks.

Finally, an issue with existing profiling tools is they provide either end-to-end throughput metrics or low-level operator traces, but this data is often fragmented and lacks explicit component-level semantics. In large systems, practitioners frequently observe performance hotspots in traces but cannot easily attribute them to specific model components. ML practitioners generally want to be able to directly attribute performance to model-level semantics to answer questions like what module is currently the biggest latency bottleneck or be able to visualize and break down MFU per module. 

To address the above challenges, we propose a unified, component-level profiling system that brings performance clarity at the module and component levels. Organizing data around these components enables user-friendly, hierarchical performance analysis.


We introduce \textit{Component Benchmark} (CB), a lightweight, extensible system for managing complex, heterogeneous profiling data that runs on a single GPU. The system follows a minimalist design, with a core of ~1000 lines of code. Leveraging PyTorch module hooks, Component Benchmark captures submodule inputs and profiles each component independently in a hierarchical manner. This approach reconstructs a tree-structured performance view of the model, allowing users to quickly localize performance hotspots within complex model architectures.

Beyond the core design, CB provides a plugin-based framework that supports extensible profiling and analysis. Users can extend CB through \textit{Component Provider} to adapt to a customized training stack, \textit{Preprocessor Plugins} to manage submodule and its input, \textit{Profiler Plugins} to collect necessary runtime information/traces, and \textit{Result Plugins} to manage and visualize final results. CB organizes and presents profiling results in an interactive, hierarchical format, enabling users to explore model performance with a significantly lesser barrier to entry and making detailed performance analysis increasingly accessible beyond performance specialists.

The scale of industrial recommendation training makes efficiency a first-order concern. These models are terabyte-scale, train continuously on thousands of GPUs, and ingest data on the order of $10^{11}$ examples per day. At this scale, even single-digit percentage improvements in throughput compound into fleet-wide capacity efficiencies. Realizing those improvements, however, requires knowing which component to optimize - a question that end-to-end throughput numbers cannot answer and operator-level traces answer at the wrong granularity. CB closes this gap by decomposing a model into its constituent sub-modules and its corresponding performance numbers. 

We discuss how Component Benchmark manages and analyzes profiling data hierarchically for representative DLRM-like models, illustrating how computation, memory, and execution time are distributed across model components. We also share our experience on how CB can be leveraged to analyze practical performance bottlenecks and was used to achieve large-scale optimizations.






\section{Related Work}

\noindent \textbf{General-purpose GPU profilers.}
Tools such as NVIDIA Nsight Systems~\cite{nsightsystems} and Kineto~\cite{kineto} provide kernel-level and operator-level GPU traces. PyTorch Profiler builds on Kineto and offers a \texttt{with\_modules} option that attributes operators to their enclosing \texttt{nn.Module}. However, this attribution is performed in-situ during a full model execution: it cannot isolate a submodule's resource consumption from cross-module effects such as kernel scheduling overlap, cache warmth, and compiler fusions. Under PyTorch 2 compilation~\cite{pytorch2}, module boundaries are further erased, making operator-to-module mapping unreliable. CB takes a different approach: it extracts each submodule with its runtime inputs and profiles it independently, preserving module-level semantics regardless of compiler transformations.

\noindent \textbf{ML performance prediction and benchmarking.}
MLPerf~\cite{mattson2020mlperf} provides standardized end-to-end benchmarks for comparing hardware and software stacks but does not decompose performance by model component. DeepSpeed~\cite{rasley2020deepspeed} includes a FLOP profiler that reports per-module FLOP counts; however, it does not provide hierarchical multi-metric analysis (latency, memory, bandwidth, MFU) or interactive visualization. CB complements these tools by offering component-level decomposition that helps practitioners understand \textit{where} within a model the bottlenecks lie, rather than only measuring aggregate throughput.

\noindent \textbf{Recommendation system profiling.}
DeepRecSys~\cite{gupta2020deeprecsys} benchmarks end-to-end inference latency for recommendation models under realistic serving conditions, focusing on system-level metrics such as tail latency and throughput under load. This is complementary to CB, which targets training-time, submodule-level profiling to guide architecture and kernel optimization. To our knowledge, CB is the first profiling system specifically designed for hierarchical, component-level performance analysis of recommendation model architectures.

\section{System Overview}

\begin{figure}[!t]
    \centering
    \includegraphics[width=0.5\textwidth]{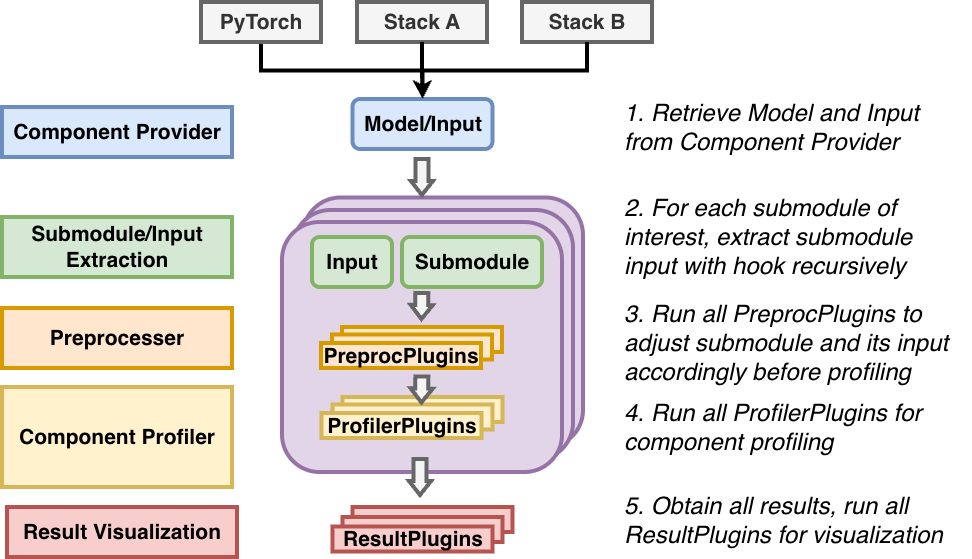}
    \caption{Component Benchmark architecture and end-to-end workflow}
    \label{fig:cb_arch}
\end{figure}


\subsection{Core Design}

Component Benchmark (CB)'s architecture and workflow is shown in Figure~\ref{fig:cb_arch}, with the following subsystems: a \textbf{Component Provider}, which adapts CB to heterogeneous training frameworks (Section~\ref{sec:plugin_design}); a \textbf{Submodule/Input Extraction} module, which uses PyTorch hooks to capture submodules and their runtime inputs (Section~\ref{sec:hook}); a \textbf{Preprocessor}, which performs data and model augmentation prior to profiling (Section~\ref{sec:plugin_design}); a \textbf{Component Profiler}, which executes component-level profiling (Section~\ref{sec:plugin_design}); and a \textbf{Result Visualizer}, which aggregates, post-processes, and presents the profiling results (Section~\ref{sec:plugin_design}).



\subsubsection{Leveraging PyTorch Hooks}
\label{sec:hook}


CB is PyTorch-native and builds on the hierarchical abstraction of PyTorch modules. Although this abstraction facilitates rapid model exploration, component-level information is frequently discarded at runtime. This loss is primarily due to the tracing mechanisms employed by modern systems, such as the PyTorch 2 Compiler~\cite{pytorch2}.



Prior to CB, we explored multiple approaches to recover module-level execution information, including injecting additional trace annotations and modifying compiler internals; however, these methods were not reliable at scale. CB instead adopts a simple and robust design: it explicitly captures each and every submodule of interest together with its runtime inputs at selected component boundaries and performs local profiling at the component level. CB implements this approach with PyTorch hooks. Given a top-level model, CB registers forward hooks on submodules of interest (by default, all submodules). During execution, these hooks capture submodule inputs, allowing CB to reconstruct fine-grained forward and backward execution behavior for each component.



\begin{table*}[t]
\centering
\caption{CB's extensible subsystems and some selected built-in functionalities}
\begin{tabularx}{\textwidth}{|l|l|X|}
\hline
\textbf{Category} & \textbf{Concrete Examples} & \textbf{Description} \\
\hline\hline
\multirow{2}{*}{Component Providers} & PyTorchProvider & Simple Provider for PyTorch model/data \\
\cline{2-3}
                                     & RecSysProvider &  Extract Model/data from RecSys stack \\
\hline
\multirow{2}{*}{Preprocessor Plugins} & Batch size adjustment & Replicate and augment submodule input \\
\cline{2-3}
                                      & Mixed precision & Apply mixed precision or low precision before profiling \\
\hline
\multirow{3}{*}{Profiler Plugins}     & Flops Counter & Calculate submodule FLOPs using PyTorch FlopCounterMode \\
\cline{2-3}
                                      & Memory snapshots & Capture PyTorch memory snapshot for submodule \\
\cline{2-3}
                                      & Trace collection & Capture Kineto traces \\
\cline{2-3}
                                      & Kernel Analysis & Aggregate and analyze perf metric at kernel level \\
\hline
\multirow{2}{*}{Result Plugins}       & Icicle view & Generate Icicle view to provide hierarchical and interactive profiling information \\
\cline{2-3}
                                      & Model graph visualization & Generate Tensorboard model graph  \\
\hline
\end{tabularx}

\label{tab:cb_plugins}
\end{table*}

\subsubsection{Recursive profiling}

CB first enumerates all submodules and orders them by hierarchical depth, from the top-level model to lower-level components, and profiles them in a breadth-first manner. Users can configure the profiling depth and apply rules to filter out uninteresting submodules. For each submodule, CB captures its runtime inputs via hooks, runs \textit{Preprocessors}, and executes the built-in and user-defined \textit{Profiler Plugins}.

If profiling a submodule fails (e.g., when a top-level module is too large to fit on a single GPU), CB skips that module and continues profiling its child submodules. This can occur in some high-level modules of very large models. In such cases, CB estimates the parent module's behavior by aggregating results from its children; this aggregation approach reconstructs module-level information hierarchically.




\begin{figure}[t]
\centering
\begin{minipage}{0.95\columnwidth}
\small
\begin{verbatim}
# Component Provider
class ComponentProviderInterface:
    def get_model_and_input(self):
        return model, input


# Preprocessor
class PreprocessorInterface:
    def preproc(self, submodule, input):
        return submodule, input


# Profiler
class ProfilerInterface:
    def profile(self, submodule, input):
        return ProfileResult


# Result Visualization
class ResultVisualizerInterface:
    def get_result(self, profiler_results):
        return HTML


# CB workflow 
model, input = provider.get_model_and_input()

for m in bfs_traverse(model):
    subinput = capture_input_with_hook(
        model, input, m)

    for preproc in preprocessor_plugins:
        m, subinput = preproc.preproc(m, subinput)

    for profiler in profiler_plugins:
        result = profiler.profile(m, subinput)
        profiler_results.append(result)

for visualizer in result_visualizer_plugins:
    visualizer.get_result(profiler_results)

\end{verbatim}
\end{minipage}
\caption{Pseudocode for the extensible Plugin interfaces in Component Benchmark subsystems and the overall workflows}
\label{fig:cb_pseudocode}
\end{figure}

\subsubsection{Plugin Design}
\label{sec:plugin_design}

Table~\ref{tab:cb_plugins} summarizes CB's extension points and some of the built-in functionalities. We adopt a simple yet extensible plugin architecture that allows users to customize and extend the system at multiple stages of the profiling pipeline. Specifically, CB supports the following extension points:

\textbf{Component Providers} integrate CB with different training frameworks by extracting the relevant model and its inputs from each specific stack.

\begin{figure*}[t]
    \centering
    \includegraphics[width=\textwidth]{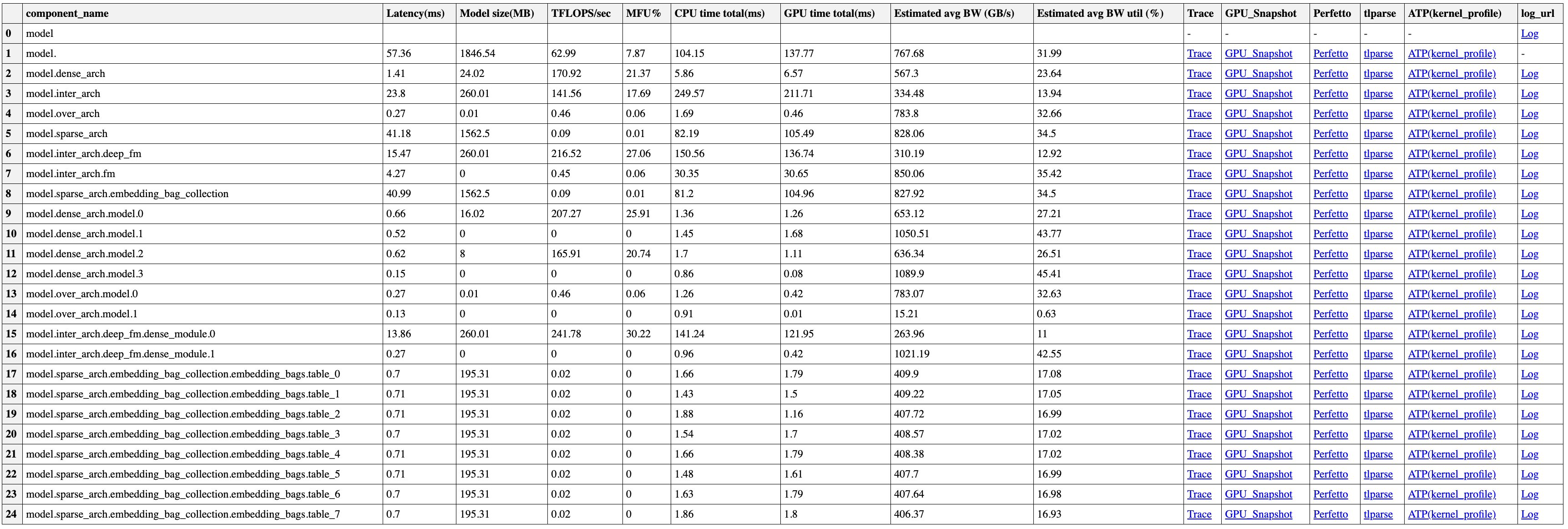}
    \caption{Benchmark Result Table for the DLRM model}
    \label{fig:table_view}
\end{figure*}

\textbf{Preprocessor Plugins} are executed before profiling each submodule and are used to augment the module/input. Typical preprocessors include batch size and mixed-precision adjustment.

\textbf{Profiler Plugins} implement the core profiling logic. Each profiler plugin is defined as a function \texttt{profile(module, input)} that produces a set of profiling artifacts. CB provides several built-in profiler plugins, including trace collection and analysis, memory snapshots, and kernel-level analysis. These plugins form the core profiling capabilities of CB and also serve as reference implementations for users to develop and contribute custom plugins.

\textbf{Result Plugins} are executed after all target submodules have been profiled and are responsible for post-processing and aggregating profiling results. Users may define custom result plugins to support application-specific analysis and visualization.

Figure~\ref{fig:cb_pseudocode} provides pseudo code of how these extensible components interact with each other.

\subsection{Components Management and Profiles Management}

\subsubsection{Components Management}

We integrated CB into recommendation training stacks by customizing \textit{Component Providers}. CB requires a top-level module and its corresponding inputs in \textit{Component Providers}; we examined the training stack and implemented the logic to retrieve the model and data. The interface is simple to follow, and we observed direct contributions.


To obtain real input, we leverage the data-loading logic embedded in the training stack. In most cases, loading a single batch is representative and sufficient for profiling. For other advanced use cases, the \textit{Component Provider} can be extended to support multiple batches, simulating dynamic input shapes encountered in the real world.

Some preprocessing is also applied to the training stack in the corresponding \textit{Component Providers}. For example, in recommendation models~\cite{naumov2019deep}, embedding table lookups before the main dense model can consume several terabytes of memory. To enable local profiling, we reduce the embedding hash size to fit within a single GPU.

CB scales down the batch size during the submodule input extraction phase, reducing memory usage. After capturing representative submodule input, CB replicates tensors to match the original batch size in \textit{Preprocessor Plugins}. This approach preserves the input distribution and effectively reduces out-of-memory (OOM) errors when profiling large models locally.

\subsubsection{Profiles Management and Visualization}

CB transforms raw profiling data into actionable insights through interactive reports. It automatically aggregates key performance metrics such as computation time, memory usage, and FLOPs, organized by model hierarchy. The reports include intuitive tables and visualizations that highlight performance hotspots, allowing users to quickly identify bottlenecks. Users can filter, explore, and export results for further analysis or sharing, which supports effective collaboration and model optimization.

\section{Experimental Analysis}

To illustrate the functionality and the effectiveness of Component Benchmark, we benchmark a common recommendation model backbone: the open source Deep FM-based~\cite{ossDeepFM} DLRM model~\cite{naumov2019deep} and a standard LLM backbone NanoGPT~\cite{karpathy2023nanogpt} on one Nvidia H100 GPU. We demonstrate how CB works hierarchically across a complex model and present it in a user-friendly way that facilitates model performance analysis. We wrap the model and data in \textit{PyTorchProvider} and invoke CB with plugins for MFU, kineto trace, memory snapshot, and our custom visualization and report generation plugins. Note that this experiment is for demonstration purposes, illustrating what CB can provide and how it benefits the performance analysis of recommendation system models.


\subsection{Benchmark Results}

\subsubsection{Result table}

The result table report, as depicted in Figure~\ref{fig:table_view}, provides a comprehensive, hierarchical view of model performance. Each row in the table corresponds to a specific submodule with the same depth, identified by its unique path. This path reflects the model's recursive structure, allowing you to trace performance metrics from the top-level model down to the leaf submodules. This recursive structure enables detailed performance analysis at every level, helping users pinpoint bottlenecks or inefficiencies within specific submodules. The following profiling information is provided with each submodule:
\begin{itemize}
    \item \textbf{Component\_name}: The hierarchical name/path of the module or submodule, showing its position within the overall model structure.
    \item \textbf{Latency (ms)}: The time taken by the module to process a batch, measured in milliseconds.
    \item \textbf{Model size (MB)}: The static memory footprint of the module, in megabytes.
    \item \textbf{TFLOPS/sec}: The throughput of the module, measured in teraFLOPS per second.
    \item \textbf{MFU\%}: Model FLOP Utilization, indicating how efficiently the module uses available compute resources.
    \item \textbf{CPU/GPU time total (ms)}: The total time spent on CPU and GPU for the module.
    \item \textbf{Estimated avg BW / BW util}: Estimated average bandwidth and its utilization.
    \item \textbf{Trace, GPU\_Snapshot, Perfetto, kernel\_profile, log\_url}: Links to various profiling and visualization tools for deeper analysis.
\end{itemize}

\subsubsection{Icicle View Visualization}

\begin{figure*}[t]
    \centering
    \includegraphics[width=\textwidth]{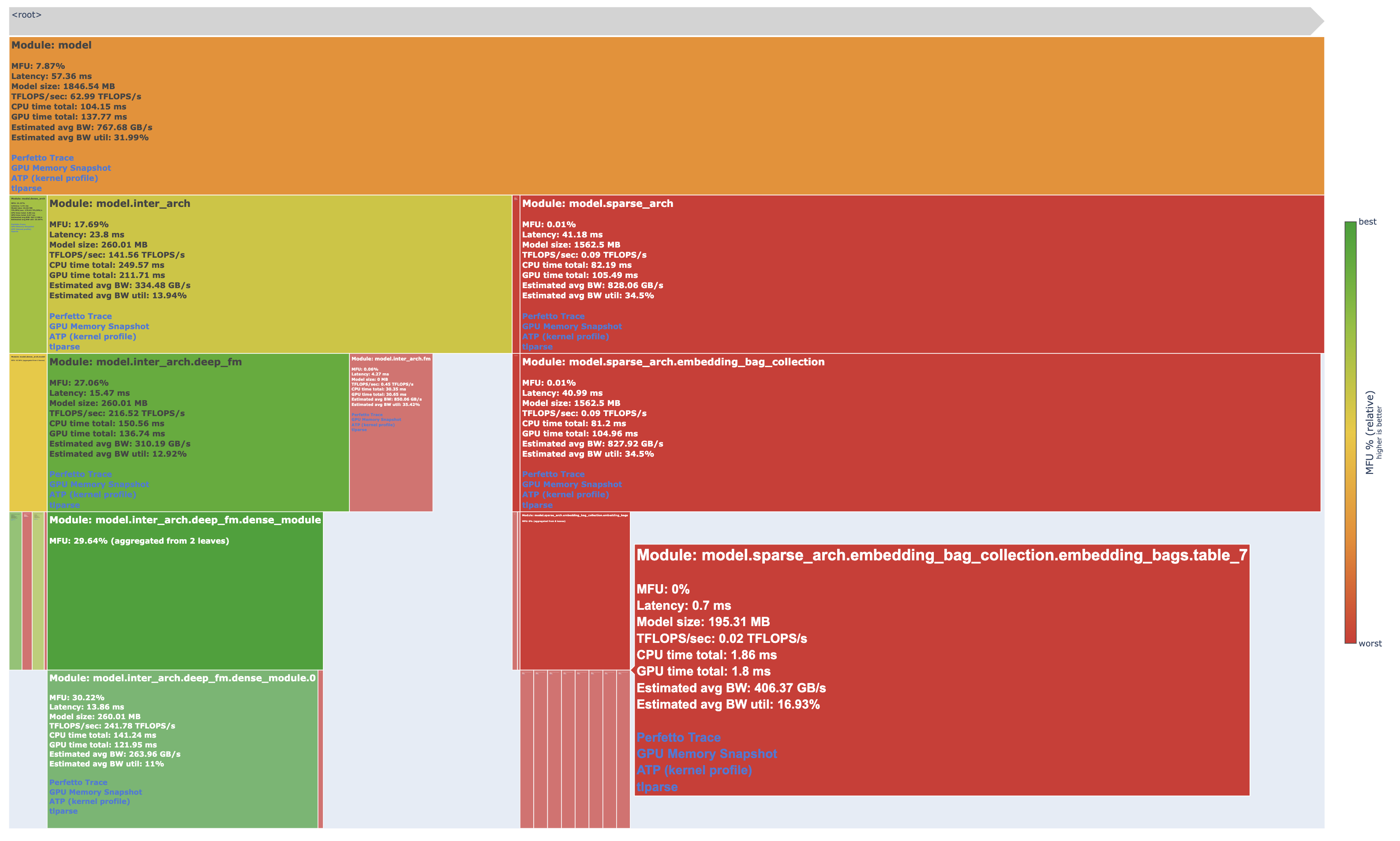}
    \caption{DLRM Benchmark Icicle Visualization. (Note: while the Icicle view font size may appear small in static figures, the interactive user interface allows flexible zooming.)}
    \label{fig:icicle_view}
\end{figure*}


The icicle view provides an interactive, hierarchical visualization of the model's component structure and performance metrics. Each node in the icicle chart represents a submodule, organized in a tree structure from the root module to leaf submodules. Figure~\ref{fig:icicle_view} shows the Icicle view benchmarking the DLRM model. Note that while the font size in the Icicle view figures may appear small in this paper, the interface supports interactive, flexible zooming.

This Icicle enables users to quickly comprehend the overall architecture: the model consists of 4 modules: sparse\_arch and dense arch, which performs lookups from embedding tables, and learns from float features respectively producing output embeddings, inter\_arch crosses them into a wide vector per example and over\_arch projects and squashes that into a probability in [0,1].

More importantly, the icicle allows us to view performance numbers per sub-module. Depending on the optimization target (we chose MFU\% here), the icicle is color coded accordingly to produce a fairly intuitive plot where the sub-modules traverse a red-green spectrum corresponding to the sub-module MFU\%. Users can interactively click on any module to access detailed performance metrics. At first glance, we are able to confirm that the sparse\_arch module is almost entirely memory bandwidth-bound with low arithmetic intensity given low FLOPs and large bandwidth utilization per module.

\begin{figure*}[t]
    \centering
    \includegraphics[width=\textwidth]{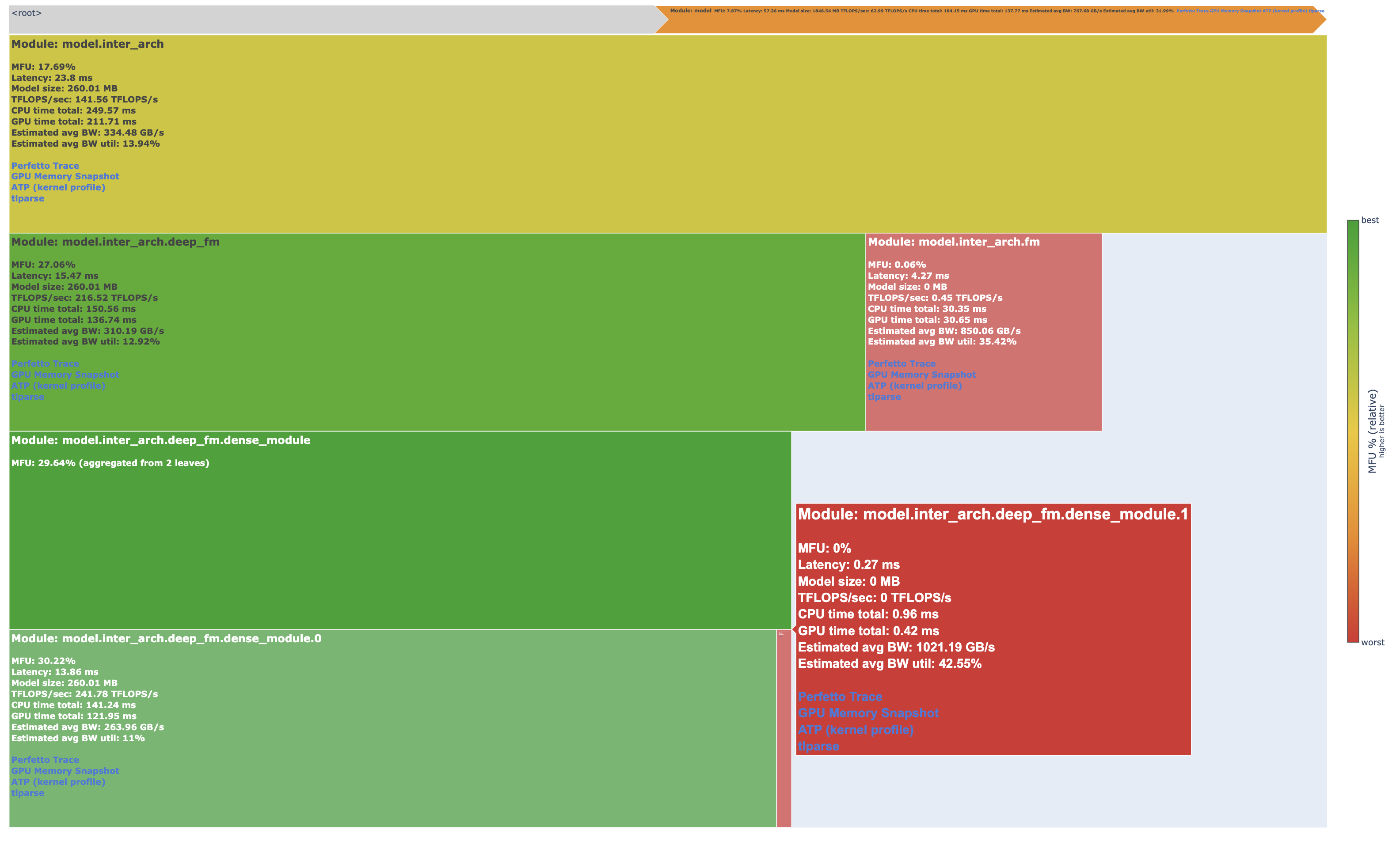}
    \caption{Zoom-in on the inter-arch module}
    \label{fig:interarch}
\end{figure*}

\begin{figure*}[t]
    \centering
    \includegraphics[width=\textwidth]{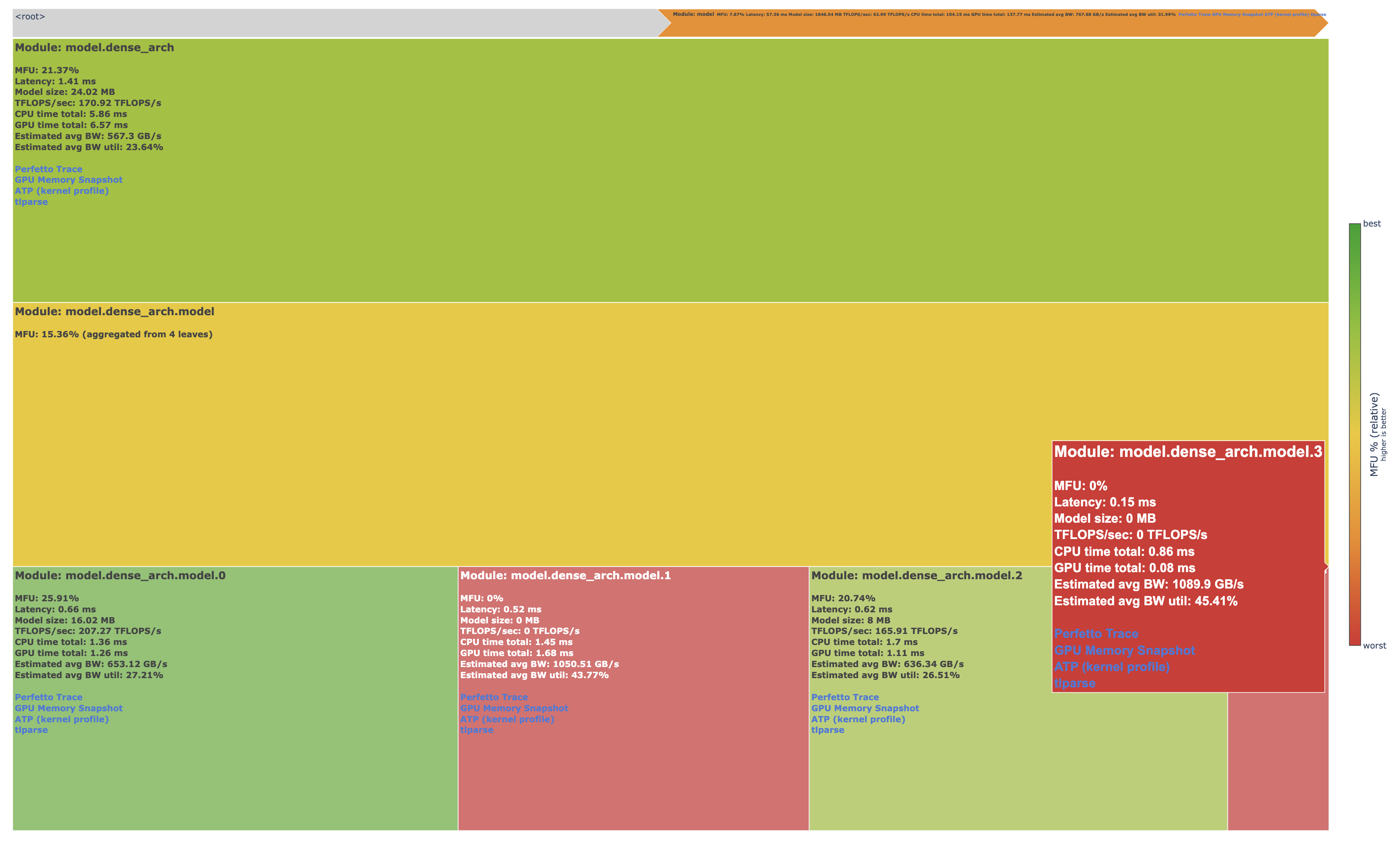}
    \caption{Zoom-in on the dense-arch module}
    \label{fig:densearch}
\end{figure*}

Diving deeper into the model, we see Figure~\ref{fig:interarch} which zooms into the inter\_arch module. We infer the following information traversing each level: 
\begin{itemize}
    \item The inter\_arch.deep\_fm and inter\_arch.fm modules are sibling sub-modules that are compute (MFU: 27.06\%) and bandwidth-bound (MFU: 0.06\%) respectively. This follows directly from the fact that the deep\_fm sub-module performs a large matrix multiplication on the sparse and dense embeddings while the fm sub-module computes the factorization-machine cross term which simplifies into 3 passes of small element-wise + reduction ops and large memory transfer.
    \item The inter\_arch.deep\_fm.dense\_module is a "container" sub-module that has no compute specific to its own module and has its performance aggregated as the latency-weighted mean over measured leaves.
    \item The sub-modules dense\_module.0 and dense\_module.1 are on opposite ends of the MFU spectrum due to module 0 being a torch.nn.Linear multiplication and module 1 being a ReLU operation which correctly reports 0 FLOPs performed.
\end{itemize}

A similar scenario is observed in Figure~\ref{fig:densearch} where we zoom into the dense\_arch module. 
\begin{itemize}
\item The dense\_arch sub-module is a single MLP stack whose leaves alternate between compute-bound torch.nn.Linear layers (dense\_module.0 at MFU: 25.91\%, dense\_module.2 at MFU: 20.74\%) and bandwidth-bound ReLU activations (dense\_module.1 and dense\_module.3, both at MFU: 0\%).
\item The two Linear layers differ from each other as well (MFU: 25.91\% vs 20.74\%), which the hierarchy attributes to shape rather than to any difference in kind: dense\_module.0 multiplies 1024x4096 while dense\_module.2 multiplies the narrower 4096x512.
\end{itemize}

\subsubsection{Extensibility to LLMs}

Another advantage of CB is that this is model agnostic and can be used to benchmark PyTorch-based models in general. CB's provider abstraction requires only two things from a workload: a top-level nn.Module and a representative input batch. Nothing in the framework is specific to recommendation architectures - the recursive walk, the input-capture mechanism, and every analysis plugin operate on the PyTorch module tree alone. To test how far that generality holds, we applied CB unchanged to NanoGPT~\cite{karpathy2023nanogpt}, an open-source decoder-only language model whose structure is the opposite of the recommendation architectures CB was designed around: a homogeneous stack of GEMM-dominated transformer blocks with no embedding-bag lookups, no jagged inputs, and no feature-interaction modules.

\begin{figure*}[t]
    \centering
    \includegraphics[width=\textwidth]{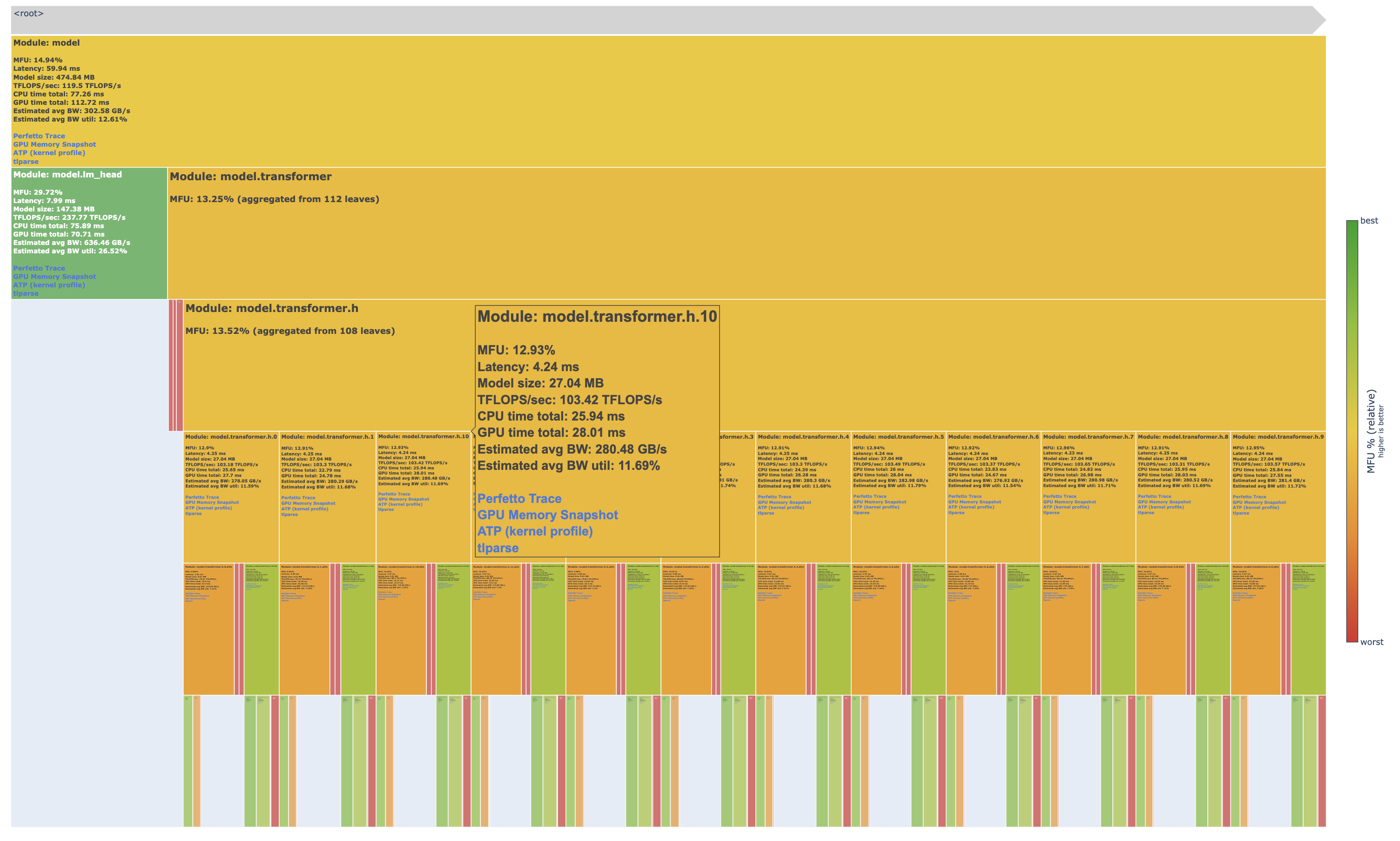}
    \caption{Top-level icicle plot of nanoGPT}
    \label{fig:nanogpt}
\end{figure*}

The top-level icicle in Fig~\ref{fig:nanogpt} shows the homogenous and repeated transformer blocks that are characteristic to large language models. We see 12 transformer blocks that account for the large majority of the model's runtime where these are indistinguishable from one another. Beside them sits lm\_head, a single Linear layer that is the model's only green cell (MFU\%), carrying a large share of total FLOPs in one leaf while occupying a much smaller share of the width - the visual signature of high arithmetic intensity. 

\begin{figure*}[t]
    \centering
    \includegraphics[width=\textwidth]{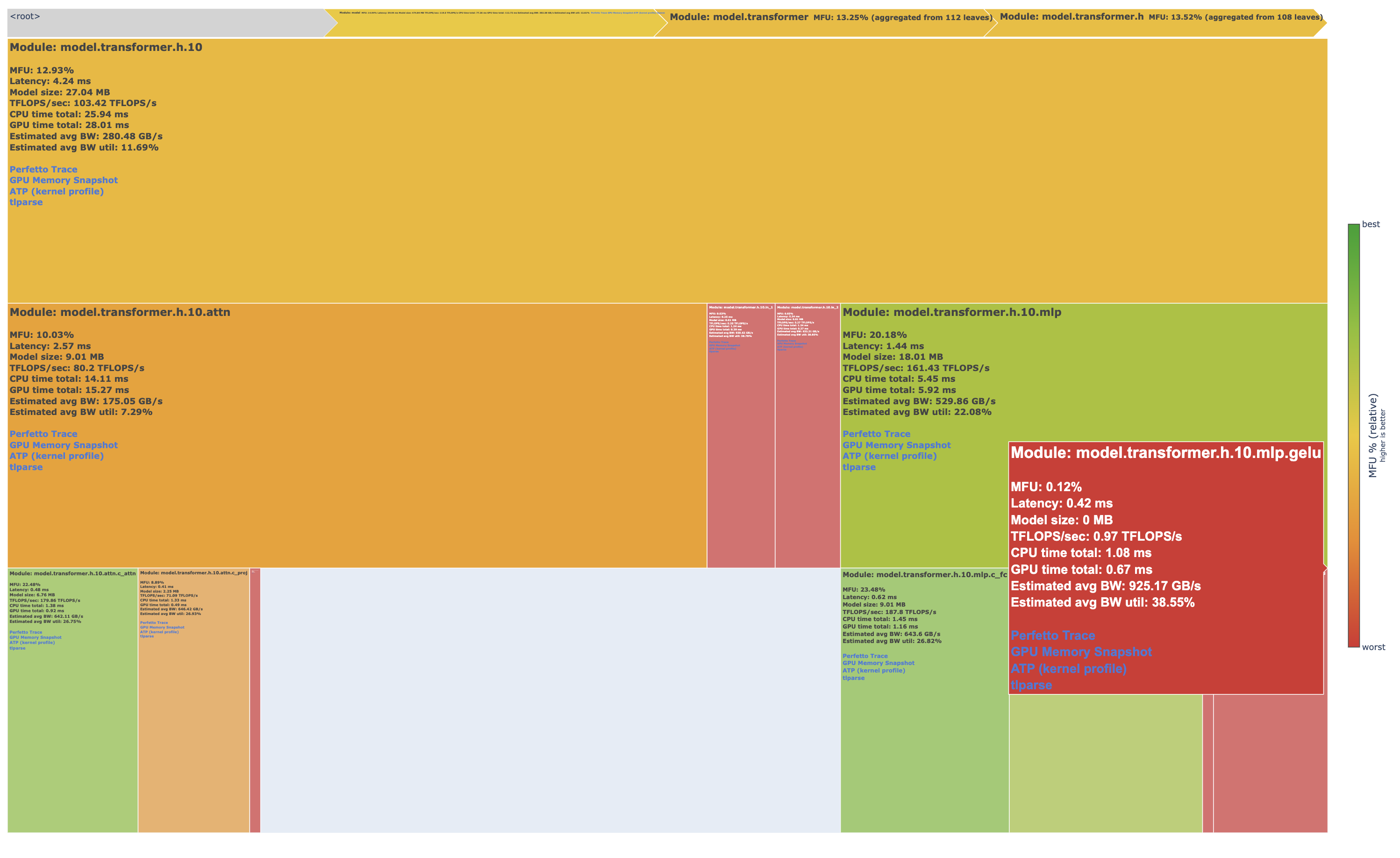}
    \caption{Zoomed-in nanoGPT transformer block}
    \label{fig:nanogpt_zoomed_in}
\end{figure*}

If we zoom in on one such transformer block in Fig ~\ref{fig:nanogpt_zoomed_in}, we make the following observations: 
\begin{itemize}
    \item The attn block has half the MFU (10\%) as the MLP block (20\%). This would naturally drive an investigation into the exact attention kernel being used which is possibly sub-optimal given the input Q,K,V dimensions. As mentioned earlier, this kernel trace is easily obtainable via the "Perfetto Trace" link that exists per sub-module and would help confirm our hypothesis.  
    \item The red blocks in the icicle are low-MFU (0-1\%) and on interactively hovering over the module, we can see this makes sense since these are layer-norm, GeLU and dropout layers which by definition have a lower number of FLOPs given the memory movement. This module-level attribution might in-turn lead to fusing these operations to increase overall MFU\% and being more compute efficient.   
\end{itemize}

These are preliminary observations on open-sourced models and are purely meant for demonstration purposes. Nonetheless, the same tooling is applied on larger models where surface-level metrics and associated profiling links 
per module enabled bottleneck elimination and reduced barrier to entry for contributors.

\section{Results in Practice}
The following case studies demonstrate the practical benefits of CB in real world usage, while also highlighting the limitations of existing benchmarking tools.\\ 

\noindent \textbf{Locally optimized kernels do not translate to globally optimal models}: CB's per-submodule MFU report isolated several low-utilization GEMV kernels in a given Attention-based module. Engineers found that re-authoring certain memory-bound GEMM/GEMV operations to instead dispatch elementwise kernels - even though those kernels are individually slower than the vendor library GEMM - enlarged the PT2 fusion region and improved memory access patterns, producing a net training and serving throughput gain of 3\%. A whole-model trace reveals that a region is slow, but cannot test the hypothesis that making a component slower makes the model faster. 

\noindent \textbf{Opportunity sizing before integration}: Only a over\_arch-based module was being compiled with PT2, leaving the benefit of extending compilation to the entire model unknown. Benchmarking the model in isolation with and without PT2 reduced iteration time from 83 ms to 50 ms, sizing the opportunity before any integration effort was incurred and justifying work that delivered 34\% QPS increase. The measurement is unreachable by whole-model methods, which cannot attribute compiler benefit to an uncompiled submodule.

\noindent \textbf{Attributing a cross-model regression}: A generative-recommender ranking model consumed 2.8× the memory and 39\% more FLOPs than its predecessor, with no attribution for either. CB's per-module memory and FLOPs breakdown localized the increase to two specific changes: a heavier event model and enlarged set-transformer blocks; restoring the baseline dimensions recovered 10\% QPS and 22\% memory savings. 

\noindent \textbf{Targeted sub-module compression}: Under a resource-reallocation mandate, tracing showed the GPU stalling on both communication and computation but gave no indication of which components to shrink. CB's joint TFLOPS-and-latency view identified two specific low-performing submodules, and after scaling down only those, the post-optimization trace shows the compute stall almost entirely eliminated. This outlines both the lack in current profiling tools that failed to answer the question and the capability of our tool that did, making it a clear demonstration of our claim.

\section{Practical Lessons Learned}

\noindent \textbf{Profiling at the right level:} While existing tools like Kineto traces provide low-level execution details, in the diverse recommendation model space, profiling at the submodule level is important to enable ML practitioners and performance engineers to identify performance bottlenecks accurately and take actionable optimization steps by correlating performance directly to each PyTorch module.

\noindent \textbf{Importance of benchmarking in large data systems:} Profiling tooling must be designed for the environment it measures. At scale it is important that performance can be attributed to a unit an engineer can act on and measured against real data. CB addresses this by reducing the unit of measurement from the job to the submodule. CB design components ensure we are able to replicate input data while being able to run performance analyses on one GPU. 

\noindent \textbf{Simple and extensible design:} We found that CB can be further leveraged for use cases we had not originally envisioned, including numeric comparison: using CB to verify submodule numeric equivalence, and hardware bring-up: benchmarking submodule performance on new hardware to identify hotspots.


\noindent \textbf{Visualization is important:} Existing profiling tools may already provide very rich information, but we found that engineers still prefer a more illustrative, intuitive user experience. Lowering the barrier for performance profiling and optimization is both important and rewarding. The hierarchical visualization, in particular, has been widely welcomed by engineers.



\section{Conclusion}


In this paper, we presented Component Benchmark (CB), a simple yet extensible system, and a recursive submodule-level profiling mechanism to address the unique complexities of recommendation model performance analysis. With its minimalist philosophy, extensible plugin architecture, robust core implementation, and user-friendly result visualization, CB brings unique perspective to the ML infrastructure community in analyzing and optimizing large scale recommendation model training. 


\end{document}